\documentclass[reprint,amsmath,amssymb,aps,prb]{revtex4-1}
\usepackage[dvipdfmx]{graphicx} 
\usepackage{dcolumn}
\usepackage{color}
\usepackage{bm}      
\usepackage{xspace}

\newcommand{\be}{\begin{equation}}
\newcommand{\ee}{\end{equation}}
\newcommand{\bea}{\begin{eqnarray}}
\newcommand{\eea}{\end{eqnarray}}
\newcommand{\beaa}{\begin{eqnarray*}}
\newcommand{\eeaa}{\end{eqnarray*}}
\newcommand{\alphaI}{{$\alpha$-(BEDT-TTF)$_2$I$_3$}\xspace}

\begin{document}

\title{
  Intervalley Tunneling and Crossover from the Positive to Negative
  Interlayer Magnetoresistance
  in Quasi-Two-Dimensional Dirac Fermion System
  with or without Mass Gap
}

\author{Takao Morinari}
 \email{morinari.takao.5s@kyoto-u.ac.jp}
 \affiliation{
   Course of Studies on Materials Science, 
  Graduate School of Human and Environmental Studies, 
  Kyoto University, Kyoto 606-8501, Japan
}

\date{\today}

\begin{abstract}
  We theoretically investigate the interlayer magnetoresistance
  in quasi-two-dimensional Dirac fermion systems,
  where the Fermi energy is at the Dirac point.
  If there is an intermediate insulating layer that
  has an overlap with the wave functions in
  the Dirac fermion layers,
  there appears a positive magnetoresistance regime
  due to the intervalley tunneling.
  We show that the interlayer magnetoresistance can be used
  to find whether Dirac fermions are massive or not  
  from the minimum in the interlayer magnetoresistance.
  As a specific system, we consider \alphaI under high pressure.
  We also discuss that one has to be careful in analyzing
  the crossover temperature from the positive
  to negative magnetoresistance.
  A simple picture is applied to the crossover
  in the zero temperature limit but it does not apply
  to the data at finite temperatures.
  We show that the ratio of the Fermi velocity to the scattering rate
  is evaluated from the zero temperature limit of the crossover temperature.
\end{abstract}

\maketitle

\section{Introduction}
\alphaI under pressure is an organic Dirac semimetal
and provides an attractive platform to investigate
the physics of the Dirac fermion state.\cite{Kobayashi2004,Katayama2006,Kajita2014}
The system has a quasi-two-dimensional structure
consisting of alternating layers of the BEDT-TTF molecules
with two-dimensional Dirac points
and an iodine layer, which is insulating.
(Here, BEDT-TTF is bis(ethylenedithio)tetrathiafulvalene.)
Remarkably, the Fermi energy is at the Dirac point
that is suitable for investigating the intrinsic properties
of the Dirac fermions.
Furthermore, it is possible to change the Coulomb interaction
from the weak to strong coupling regimes.
The quantum phase transition is investigated
in the strong coupling regime,
and it turns out that the system approaches
the quantum critical point without creating any mass gap
from the weak coupling regime.\cite{Unozawa2020,Kobayashi2021}

The presence of the Dirac fermion spectrum is clearly demonstrated
in the interlayer magnetoresistance.\cite{Tajima2009}
Since the Fermi energy is at the Dirac point,
the density of states vanishes at the Fermi energy in the absence of the magnetic field.
By applying a magnetic field, there appears the zero energy Landau level
that is absent in the conventional Landau levels of non-relativistic
electrons.
Because of the huge degeneracy in the zero-energy Landau level,
the interlayer magnetoresistance is expected to be negative and
be in inversely proportional to the applied magnetic field.
The experiment is consistent with the theoretical result\cite{Osada2008}
based on the zero energy Landau level.
However, the system first exhibits a positive magnetoresistance
as the applied magnetic field is increased,
and then it shows a negative magnetoresistance.
The magnetoresistance can be positive if
there is mixing between different Landau levels.\cite{Morinari10a}
Although such mixing is possible when the direction of the interlayer hopping
is nonvertical, the mechanism is not so convincing because
the spatial translation of the Landau level wave functions
upon interlayer hopping is much shorter than the magnetic length.

In this paper, we reconsider the mechanism of the positive
magnetoresistance.
We also consider how the interlayer resistivity
is used to distinguish massless and massive Dirac fermions.
We point out that there is an inter-valley interlayer hopping
due to the overlaps between the molecular wave functions in BEDT-TTF layers
and those in the iodine layers.
We show that the inter-valley hopping clearly leads to the positive magnetoresistance
under weak fields.
We also investigate the effect of the mass gap in the Dirac fermion spectrum
on the interlayer magnetoresistance.
Recently, the effect of the spin-orbit coupling was
considered in the related organic compounds.\cite{Winter2017}
A possibility of the topological insulating state
in \alphaI has been proposed.\cite{Osada2018}
We show that the interlayer magnetoresistance
can be used to clarify whether there is a mass gap or not.
We also describe a method to extract the ratio of the Fermi velocity
to the scattering rate
from the interlayer magnetoresistance measurements.

The rest of the paper is organized as follows.
In Sect.~\ref{sec:interlayer_hopping} we formulate the inter-layer hopping
with including the presence of the iodine layers,
and show that there is an inter-valley interlayer tunneling.
In Sect.~\ref{sec:interlayer_MR_formula} we derive the formula for the
interlayer conductivity under magnetic fields with and without the mass gap.
In Sect.~\ref{sec:results} we present the result
and discuss how one can distinguish the massless and massive cases.
We also discuss how the interlayer magnetoresistance experiment
is used to estimate the ratio of the Fermi velocity to the scattering rate
of the Dirac fermions.
Section~\ref{sec:conclusion} is devoted to conclusion.

\section{The model for the interlayer tunneling}
\label{sec:interlayer_hopping}
In this section we establish that there is an effective inter-valley
electron hopping between BEDT-TTF layers
by explicitly taking into account the presence
of the iodine layers in between.
As revealed by the first principles calculation,\cite{Kino2006}
there is mixing between the BEDT-TTF and I$_3^-$ molecules.
Therefore, electrons do not hop directly between adjacent BEDT-TTF layers.
They hop from a BEDT-TTF layer to the adjacent iodine layer,
and then hop to another BEDT-TTF layer.
Because of this intermediate process, the inter-valley hopping is possible.
If we denote the hopping amplitude between BEDT-TTF and iodine layers by $t'$,
then we obtain the effective hopping $t_c$ between adjacent BEDT-TTF layers
by applying the second order perturbation theory
with respect to $t'$.
The result is,
\be
  {t_c} =  - \frac{{2{{t'}^2}}}{{{U_p} + {\varepsilon _p}}} - \frac{{{{t'}^2}}}{{{U_p} - U + {\varepsilon _p}}}
  \label{eq:t_c}
  \ee
where $U_p$ and $U$ are the onsite Coulomb repulsions
at the iodine and BEDT-TTF layers, respectively.
$\varepsilon _p$ is the energy of the single-electron at the iodine layer.
The LUMO+1 and LUMO+2, where LUMO is the lowest-unoccupied-molecular-orbital,
are around 1eV above the Fermi energy and consist of I$_3^-$ molecules.\cite{Kino2006}
So, $\varepsilon _p \sim 1$eV.

We emphasize that $t_c$ is not the direct hopping amplitude of electrons
between BEDT-TTF layers.
If $t_c$ were the direct hopping amplitude,
then there would be no tunneling between different valleys
in the Dirac fermion phase.
Since $t_c$ is related to $t'$ that is
the hopping amplitude between BEDT-TTF and I$_3^-$ molecules,
the tunneling is not restricted to the same valleys
but there is also the inter-valley tunneling.

Now we present the effective interlayer tunneling term between BEDT-TTF layers
in the Dirac fermion phase of \alphaI 
that is used to derive the formula for the interlayer conductivity
under magnetic field.
From the consideration above, the inter-layer hopping
includes the inter-valley hopping.
Our model Hamiltonian for the interlayer tunneling is given by
\be
  {H_c} = {t_c}\sum\limits_{n,n',\tau ,\tau ',k,\sigma ,\ell }
  {\left( {{M_{n\tau ,n'\tau '}}c_{n\tau k\sigma ,\ell  + 1}^\dag
      {c_{n'\tau 'k\sigma ,\ell }} + H.c.} \right)}
  \label{eq:H_c}
  \ee
  Here, the operator ${c_{n\tau k\sigma ,\ell}^\dag }$ is the creation operator
  for a Dirac fermion with the Landau level index $n$, valley index $\tau$,
  the wave number $k$, and spin $\sigma$ in the $\ell$-th layer.
  A summary of the Dirac fermion Landau levels is given
  in Appendix \ref{app:DiracFermionLL}.
  We note that $M_{n\tau ,n'\tau '}$ is a function of $k$, $k'$, 
  $\tau$, and $\tau'$, in general but here we assume $k'=k$ for simplicity.
  $M_{n\tau ,n'\tau '}$ are the matrix elements
  between the Landau level $n$ at valley $\tau$
  and the Landau level $n'$ at valley $\tau'$
  that is given by
  \be
    {M_{n\tau ,n'\tau '}} = \left\langle {{n,\tau }}
 \mathrel{\left | {\vphantom {{n,\tau } {n',\tau '}}}
 \right. \kern-\nulldelimiterspace}
         {{n',\tau '}} \right\rangle,
\ee
where $\left| {n,\tau } \right\rangle$ is defined
in Appendix \ref{app:DiracFermionLL}.

\section{Interlayer magnetoresistance formula}             
\label{sec:interlayer_MR_formula}
In this section we derive the formula for the interlayer conductivity
based on Eq.~(\ref{eq:H_c}).
We assume that $t_c$ is much smaller than the intralayer scattering rate, $\Gamma$.
Under this condition the interlayer conductivity is proportional
to the tunneling rate between two adjacent layers\cite{McKenzie1998,Moses1999}
labeled by layer-1 and -2.
The tunneling Hamiltonian between these two layers is given by
  \be
    {H_{12}} = {t_c}\sum\limits_{\alpha ,\alpha '} {\left( {{M_{\alpha '\alpha }}c_{\alpha '2}^\dag {c_{\alpha 1}} + {M_{\alpha \alpha '}}c_{\alpha 1}^\dag {c_{\alpha '2}}} \right)},
    \ee
    where we denote a set of parameters, $n,\tau ,k,\sigma$
  by a single Greek symbol, $\alpha$,
  to simplify the notation.
  Since the spin is conserved upon the interlayer tunneling,
  we set
    \be
      {M_{\alpha \alpha '}} = {\delta _{k,k'}}
      {\delta _{\sigma ,\sigma '}}{M_{n\tau ,n'\tau '}}.
      \ee

The interlayer current $I$ produced by an applied voltage $V$ across the layers
is calculated as in a metal-insulator-metal junction.\cite{MahanChap9_3}
The current $I$ is associated with the change of
the number of particles in layer-1,
\be
  {N_1} = \sum\limits_\alpha  {c_{\alpha 1}^\dag {c_{\alpha 1}}}.
  \ee
  Under the applied voltage $V$,
  the chemical potential difference between the layers is
  ${\mu _1} - {\mu _2} = eV$,
  where $\mu_1$ and $\mu_2$ denote
  the chemical potentials at each layer.
  Applying the Kubo formula, we obtain
  \bea
  I\left( t \right) &=&
  - \frac{{et_c^2}}{{{\hbar ^2}}}\int_{ - \infty }^\infty  {dt'}
  \theta \left( {t - t'} \right)
  \left[
    {e^{ - \frac{i}{\hbar }\left( {{\mu _2} - {\mu _1}} \right)
        \left( {t - t'} \right)}} \right. \nonumber \\
    & & \left. \times
    \left\langle {\left[ {A\left( t \right),{A^\dag }\left( {t'} \right)} \right]} \right\rangle \right. \nonumber \\
    & & \left.
    - {e^{ + \frac{i}{\hbar }\left( {{\mu _2} - {\mu _1}} \right)\left( {t - t'} \right)}}
    \left\langle {\left[ {{A^\dag }\left( t \right),A\left( {t'} \right)}
        \right]} \right\rangle
    \right],
\label{eq:I_formula1}
\eea
where
\be
A = \sum\limits_{\alpha ,\alpha '} {{M_{\alpha '\alpha }}c_{\alpha '2}^\dag
  {c_{\alpha 1}}},
\ee
\be
  {A^\dag } = \sum\limits_{\alpha ,\alpha '} {{M_{\alpha \alpha '}}c_{\alpha 1}^\dag {c_{\alpha '2}}},
  \ee
  and
  \be
  A\left( t \right) = \exp \left( {\frac{i}{\hbar }K t} \right)
  A\exp \left( { - \frac{i}{\hbar }K t} \right),
  \ee
  with $K = \sum\limits_{\alpha ,j = 1,2} {\left( {{\varepsilon _\alpha } - \mu } \right)c_{\alpha j}^\dag {c_{\alpha j}}}$.
  Here, $\mu$ is the chemical potential in the absence of the voltage drop.
  $\varepsilon _\alpha$ denotes the energy of the single-particle state, $\alpha$,
  whose Landau energy part is given in Appendix \ref{app:DiracFermionLL}
  and the Zeeman energy, $\mu_{\rm B} B$, is included.
  (The spin $g$-factor is set to 2.)
  
  Introducing the retarded and advanced Green's functions,
  \be
     {D_R}\left( {t - t'} \right) =  - \frac{i}{\hbar }\theta \left( {t - t'} \right)\left\langle {\left[ {A\left( t \right),{A^\dag }\left( {t'} \right)} \right]} \right\rangle,
     \ee
     \be
        {D_A}\left( {t - t'} \right) =  + \frac{i}{\hbar }\theta \left( {t' - t} \right)\left\langle {\left[ {A\left( t \right),{A^\dag }\left( {t'} \right)} \right]} \right\rangle,
        \ee
and their Fourier transforms $D_R(\omega)$ and $D_A(\omega)$,
Eq.~(\ref{eq:I_formula1}) is rewritten as
\be
I = \frac{{ie}}{\hbar }t_c^2\left[ {{D_R}
      \left( {eV/\hbar } \right) - {D_A}\left( {eV/\hbar } \right)} \right].
\label{eq:formula_I2}
\ee
${D_R}\left( \omega  \right)$ and
${D_A}\left( \omega  \right)$ are obtained
from the Matsubara Green's function ${D_M}\left( i\omega_n  \right)$
by the analytic continuation.
${D_M}\left( i\omega_n  \right)$ is given by
\be
   {D_M}\left( {i{\Omega _n}} \right) = \frac{1}{\beta }\sum\limits_{\alpha ,\alpha '} {{{\left| {{M_{\alpha '\alpha }}} \right|}^2}} \sum\limits_{i{\omega _n}} {{G_{\alpha 1}}\left( {i{\omega _n} + i{\Omega _n}} \right){G_{\alpha 2}}\left( {i{\omega _n}} \right)},
   \ee
   where $\beta = 1/(k_{\rm B} T)$ with $k_{\rm B}$ the Boltzmann constant.
   $\omega_n$ and $\Omega_n$ are
   fermionic and bosonic Matsubara frequencies, respectively.
The single-body Green's function is given by
\be
  {G_{\alpha j}}\left( {i{\omega _n}} \right) = \int_0^\beta  {d\tau } {e^{i{\omega _n}\tau }}{G_{\alpha j}}\left( \tau  \right),
  \ee
with
\be
  {G_{\alpha j}}\left( \tau  \right) =  - \left\langle {{T_\tau }{c_{\alpha j}}\left( \tau  \right)c_{\alpha j}^\dag \left( 0 \right)} \right\rangle.
  \ee
  Here, $T_{\tau}$ is the imaginary-time ordering operator
  and
  ${c_{\alpha j}}\left( \tau  \right) = \exp \left( {\tau K} \right){c_{\alpha j}}\exp \left( { - \tau K} \right)$.

Introducing the spectral representation of the single-particle Green's function,
we carry out the summation over $i\omega_n$.
After the analytic continuation, we obtain
\bea
    {\mathop{\rm Im}\nolimits} {D_R}\left( \omega  \right) &=& \pi \sum\limits_{\alpha ,\alpha '} {{{\left| {{M_{\alpha '\alpha }}} \right|}^2}} \int_{ - \infty }^\infty  {d{\omega _1}} \int_{ - \infty }^\infty  {d{\omega _2}} \nonumber \\
    & & \times \left[ {f\left( {{\omega _1}} \right) - f\left( {{\omega _2}} \right)} \right]\delta \left( {\hbar \omega  - {\omega _1} + {\omega _2}} \right)
    \nonumber \\
    & & \times {A_{\alpha 1}}\left( {{\omega _1}} \right)
              {A_{\alpha' 2}}\left( {{\omega _2}} \right),
              \eea
              with $f\left( \varepsilon  \right)
              = 1/\left( {\exp \left( {\beta \varepsilon } \right) + 1} \right)$
              the Fermi-Dirac distribution function.
Here, ${A_{\alpha j}}\left( \omega  \right)$ is given by
\be
   {A_{\alpha j}}\left( \omega  \right) =  - \frac{1}{\pi }{\mathop{\rm Im}\nolimits} {G_{\alpha j}}\left( {\omega  + i\delta } \right).
   \ee
From Eq.~(\ref{eq:formula_I2}),
the interlayer conductivity is
\bea
    {\sigma _{zz}}
    &=& \frac{{2\pi {e^2}t_c^2{a_c}}}{{\hbar {L^2}}}\sum\limits_{\alpha ,\alpha '} {{{\left| {{M_{\alpha '\alpha }}} \right|}^2}} \int_{ - \infty }^\infty  {d\omega } \left( { - \frac{{\partial f}}{{\partial \omega }}} \right)
    \nonumber \\
    & & \times
    {A_{\alpha 1}}\left( \omega  \right){A_{\alpha' 2}}\left( \omega  \right)
   \label{eq:s_zz_general}
\eea
with $a_c$ the $c$-axis lattice constant
and $L^2$ the area of the layer.
Carrying out the $k$-summation, we obtain
\bea
    {\sigma _{zz}} &=& \frac{{{e^3}t_c^2{a_c}}}{{{\hbar ^2}}}B\sum\limits_{n,n',\tau ,\tau ',\sigma } {{{\left| {{M_{n\tau ,n'\tau '}}} \right|}^2}} \int_{ - \infty }^\infty
    {d\omega } \nonumber \\
    & & \times \left( { - \frac{{\partial f}}{{\partial \omega }}} \right){A_{n\tau \sigma}}\left( \omega  \right){A_{n'\tau '\sigma}}\left( \omega  \right).
    \label{eq:s_zz_formula}
\eea    
Note that we omit the indices for layer-1 and -2 hereafter
because the spectral function $A_{n\tau \sigma}$ has the same functional form.
The explicit forms of $M_{n\tau ,n'\tau '}$
are given in Appendix \ref{app:M}.

A convenient form of Eq.~(\ref{eq:s_zz_formula}) is obtained
by dividing it by
${\sigma _0} = \left( {{e^2}/\hbar {a_c}} \right){\left( {{t_c}/\Gamma } \right)^2}$.
That is,
\bea
\frac{{{\sigma _{zz}}}}{{{\sigma _0}}} &=& \frac{e}{\hbar }a_c^2B \times {\Gamma ^2}\sum\limits_{n,n',\tau ,\tau ',\sigma } {{{\left| {{M_{n\tau ,n'\tau '}}} \right|}^2}}
\nonumber \\
& & \times \int_{ - \infty }^\infty  {d\omega } \left( { - \frac{{\partial f}}{{\partial \omega }}} \right){A_{n\tau \sigma 1}}\left( \omega  \right){A_{n'\tau '\sigma 2}}\left( \omega  \right).
\label{eq:s_zz_formula_B}
\eea
If we measure the magnetic field $B$ in units of Tesla
and taking ${a_c} = 1.7488 \times {10^{ - 9}}\,{\rm{m}}$,\cite{Bender1984}
we obtain
$ea_c^2B/\hbar = 4.646 \times {10^{ - 3}} \times B\left[ {\rm{T}} \right]$.

\section{Results}
\label{sec:results}
\subsection{Massless Dirac fermions}
\label{subsec:massless_case}
\begin{figure}[tbp]
  \includegraphics[width=\linewidth]{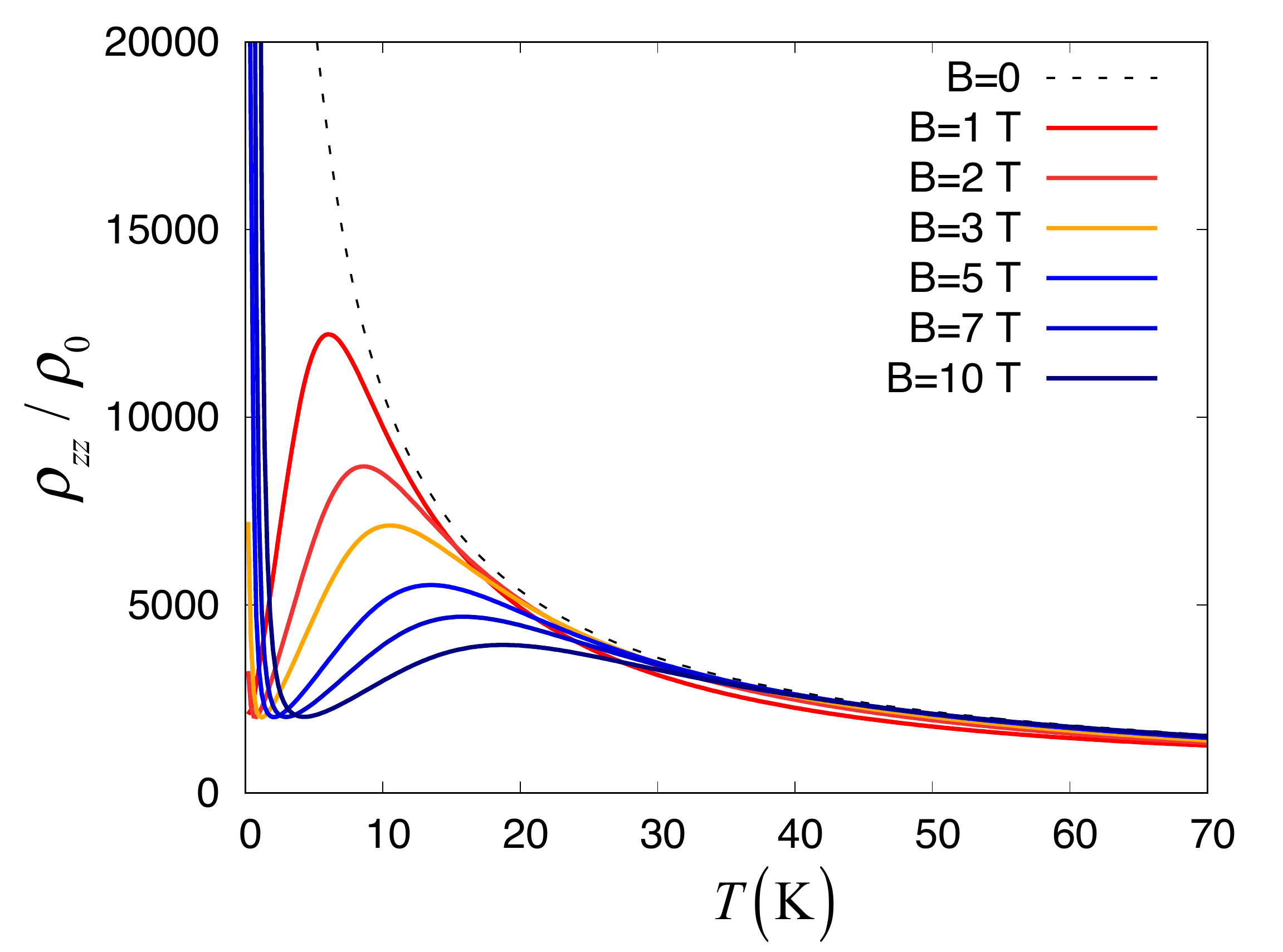}
  \caption{
    \label{fig:rzz_for_Bs}
    (Color online)     
    The interlayer resistivity as a function of temperature
    for different magnetic fields
    with $v_F=5 \times 10^4$~m/s and $\Gamma=1$ K.
    The dashed line is the interlayer resistivity at $B=0$.
    Here, ${\rho _0} = 1/\sigma_0$.
    There is crossover from
    the positive magnetoresistance at low temperatures to
    the negative magnetoresistance at high temperatures.
    The Landau-level sum is taken over $-n_{\rm max} \leq  n \leq n_{\rm max}$
    with $n_{\rm max}=300$.
  }
\end{figure}

Now we show the results for the massless Dirac fermion case.
As will be shown below we clearly see the crossover from the positive
to negative magnetoresistance regimes.
Figure \ref{fig:rzz_for_Bs} shows
the temperature dependence of the interlayer resistivity for different magnetic fields.
Hereafter, we take $v_F=5 \times 10^4$~m/s and $\Gamma=1$ K unless otherwise stated.
We plot the interlayer resistivity at $B=0$ as well,
whose formula is given in Appendix \ref{app:B_0formula}.
The steep rise close to zero temperature is associated with
the energy gap created by the Zeeman energy,
which is of no importance.
Apart from that peak,
we clearly see the crossover from the positive
to negative magnetoresistance regimes.
The interlayer resistivity shows a peak at the crossover temperature,
$T_{\rm max}$.
For example $T_{\rm max} \sim 10$~K for $B=3$~T.
In the positive magnetoresistance regime, $T<T_{\rm max}$,
the inter-Landau level mixing occurs,
and this dominates over the increase of the conductivity
due to the Landau level degeneracy,
which is proportional $B$.
This Landau level mixing arises from the intervalley scattering
upon the interlayer tunneling.
We note that at high-temperatures
the resistivity approaches to the $B=0$ value.
In order to confirm this behavior,
we need to carry out the Landau-level sum with
a sufficient number of Landau levels
because of the factor associated with
the derivative of the Fermi-Dirac function
that has a broad peak at high temperatures.

Here, we comment on the temperature dependence of $\Gamma$.
In this paper we assume a constant $\Gamma$.
In the experiment reported in Ref.~\onlinecite{Sugawara2010mr},
the temperature dependence
of the interlayer magnetoresistance was shown for different magnetic fields.
It is possible to compute the interlayer magnetoresistance theoretically
from the results above.
However, the temperature dependence of the interlayer resistivity at $B=0$
in Ref.~\onlinecite{Sugawara2010mr}
shows a broad dip that is presumably associated with the temperature dependence
of $\Gamma$.
The theoretical results above assumed a constant $\Gamma$.
So, we need to include its temperature dependence
to quantitatively explain the experiment.
This point will be investigated in the future publication.

\begin{figure}[tbp]
  \includegraphics[width=\linewidth]{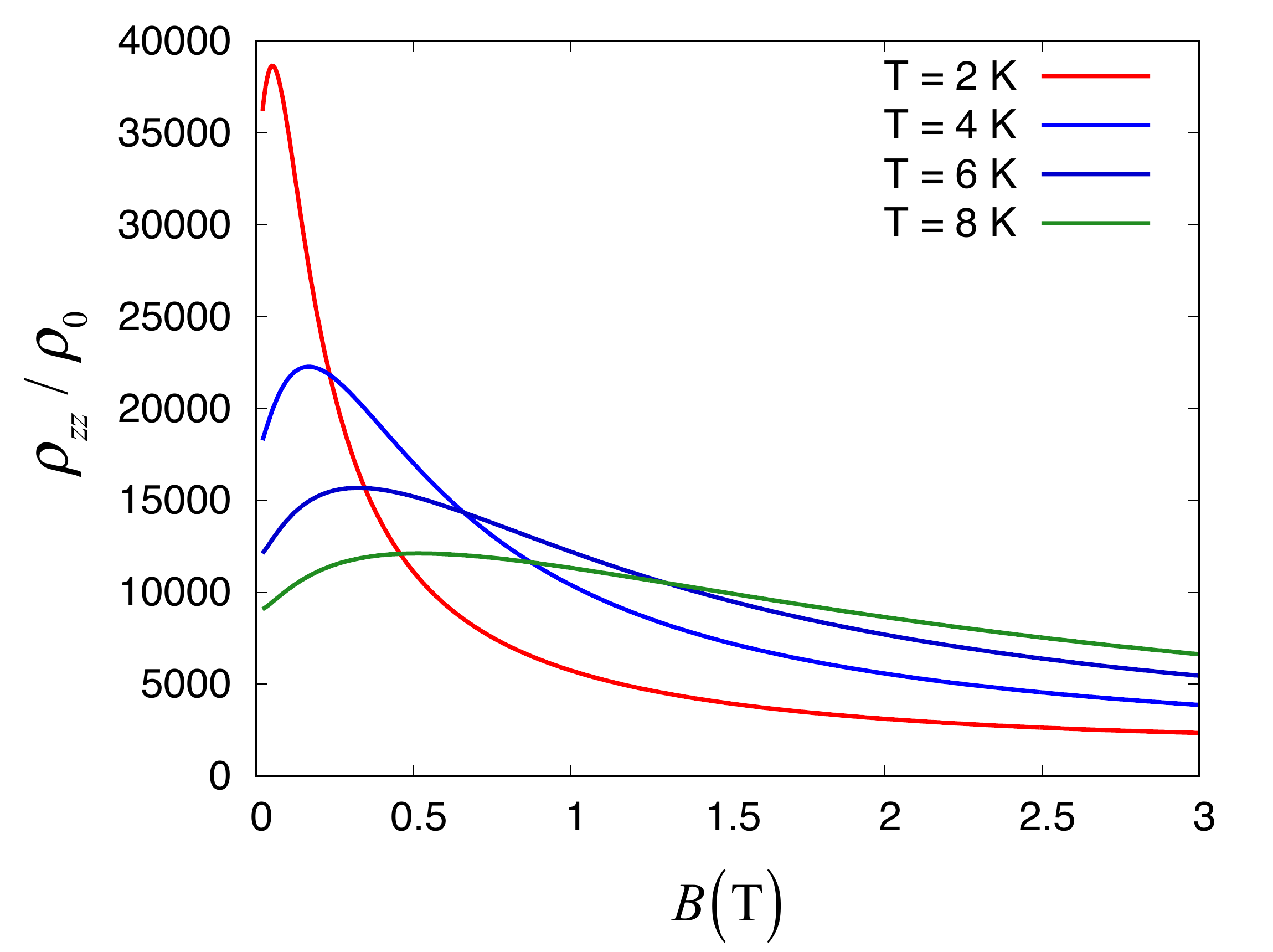}
  \caption{
    \label{fig:rzz_for_Ts}
    (Color online)
    The interlayer resistivity as a function of $B$
    for different temperatures.
    There is crossover from
    the positive magnetoresistance at low fields to
    the negative magnetoresistance at high fields.
    The physical origin of this crossover is the same as that
    observed in the temperature dependence
    of $\rho_{zz}$ (Fig.~\ref{fig:rzz_for_Bs}).
  }
\end{figure}

Before analyzing the magnetic field dependence of $T_{\rm max}$,
we discuss the magnetic field dependence
of the interlayer resistivity for different temperatures
shown in Fig.~\ref{fig:rzz_for_Ts}.
We also find the crossover from the positive 
to negative magnetoresistance regimes.
There is a peak associated with this behavior.
We would like to emphasize that the peaks appear
both in Figs.~\ref{fig:rzz_for_Bs} and \ref{fig:rzz_for_Ts}.
This is because the physical origin of the peaks are the same.
It was argued in Ref.~\onlinecite{Yoshimura2021}
that the positive magnetoresistance was possible
if the Dirac fermions were massive.
However, the positive magnetoresistance region is absent
in the magnetic field dependence of the interlayer resistivity
if one neglects the effect of the mixing between the Landau levels.
Since the origin of the positive magnetoresistance is the same,
just assuming the mass gap for the Dirac fermion spectrum
is not enough to explain the experiment.

\begin{figure}[tbp]
  \includegraphics[width=\linewidth]{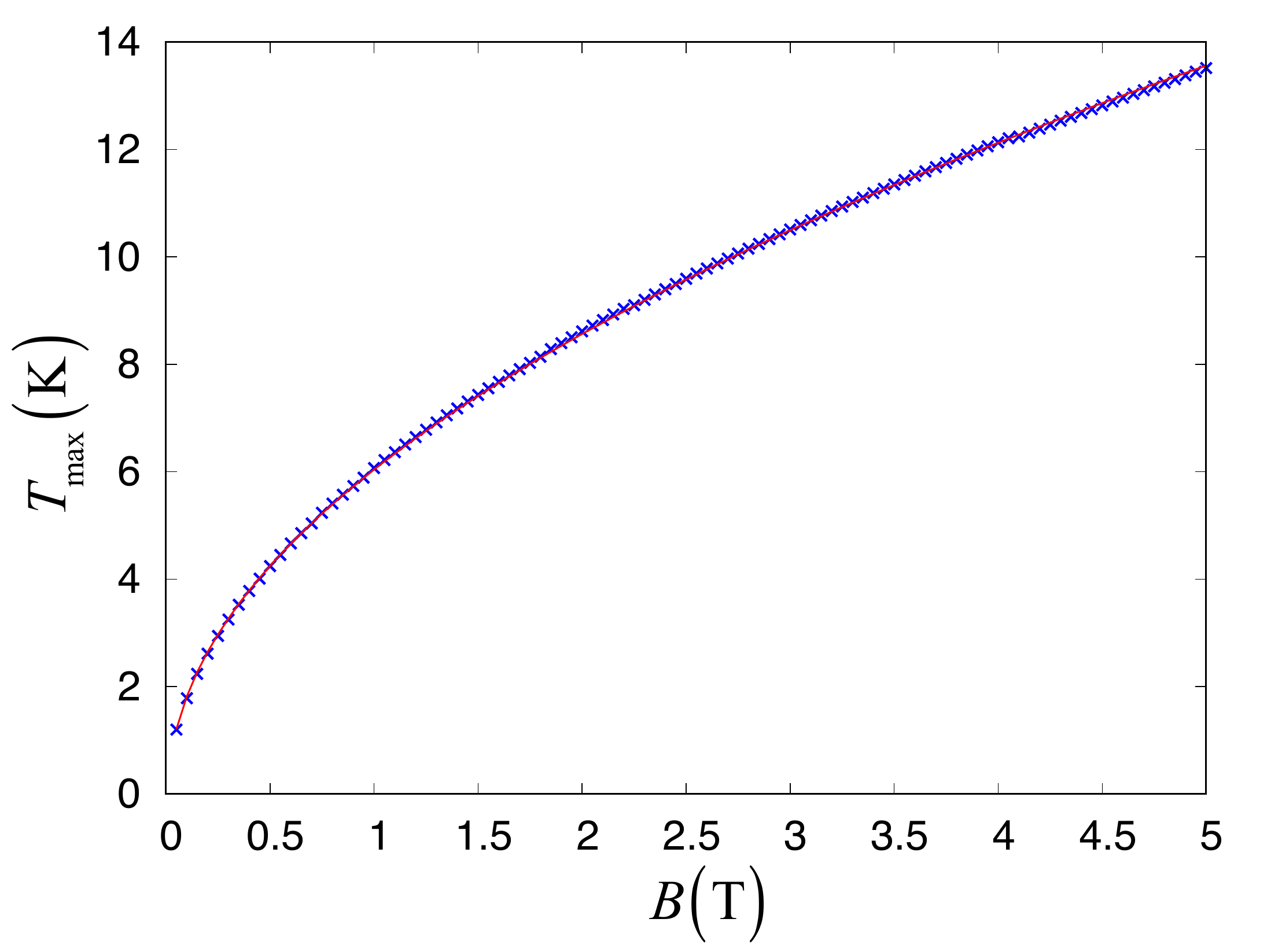}
  \caption{
    \label{fig:Tmax_m0}
    (Color online)
    The crosses show $T_{\rm max}$ as a function of $B$.
    The solid line is a fit to a phenomenological formula,
    ${T_{\max }} = C\sqrt {B - {B_0}}$,
    where $C$ and $B_0$ are fit parameters.
    From the fit we find $C=6.07$ K T$^{-1/2}$ and $B_0=1.0 \times 10^{-2}$ T.
    Note that we need more careful analysis to estimate the $B_0$ value
    as discussed in the main text.
  }
\end{figure}

Now we discuss $T_{\rm max}$ for different magnetic fields,
and we resolve some confusing situation about its interpretation.
Figure~\ref{fig:Tmax_m0} shows the magnetic field dependence
of $T_{\rm max}$.
Roughly speaking $T_{\rm max}$ appears to be proportional to $\sqrt{B}$.
One can understand this behavior as the result of the Landau level mixing.
Since the Landau levels are not evenly spaced for the Dirac fermion system
and the largest energy gap is that between the zero and first Landau levels,
the Landau level mixing occurs when this gap is equal to $\Gamma$.
If the Landau level broadening associated with $\Gamma$ is
larger than the temperature effect,
we may expect $T_{\rm max} \propto \sqrt{B}$ because
the first Landau level energy is proportional to $\sqrt{B}$.
However, one must be careful that this interpretation
applies to the temperature range, $T\ll \Gamma/k_{\rm B}$.
At finite temperatures it gives us just a qualitative picture.
The point is that the temperature effect 
is not simple because that comes from the integration of the function
multiplied by the derivative of the Fermi-Dirac function
over the whole range of the Landau levels.
In Ref.~\onlinecite{Sugawara2010mr},
$T_{\rm max}$ is analyzed using a formula
${k_{\rm{B}}}{T_{\max }}
  = c\left( {v\sqrt {2e\hbar B}  - g{\mu _B}B} \right)$,
    with $c$ and $v$ being fit parameters.
    $g=2$ is the $g$-factor and $\mu_B$ is the Bohr magneton.
  The experiment was in good agreement
  with this formula
  with $c=0.93$
  and $v=2.4\times 10^4$ m/s.
  However, this result requires close attention.
  Although $c\sim 1$,
  the value of $v$, which must correspond to the Fermi velocity,
  is almost one-half of the Fermi velocity determined
  from the Shubnikov-de Haas oscillation.\cite{Unozawa2020}
  Therefore, the simple picture above is not applicable.
  Meanwhile,
  it was argued that the Dirac fermions were massive
in Ref.~\onlinecite{Yoshimura2021}
based on a similar analysis.
But from the observation above one is not able to
conclude whether the Dirac fermions are massive or not
from the finite temperature data.

\begin{figure}[tbp]
    \includegraphics[width=\linewidth]{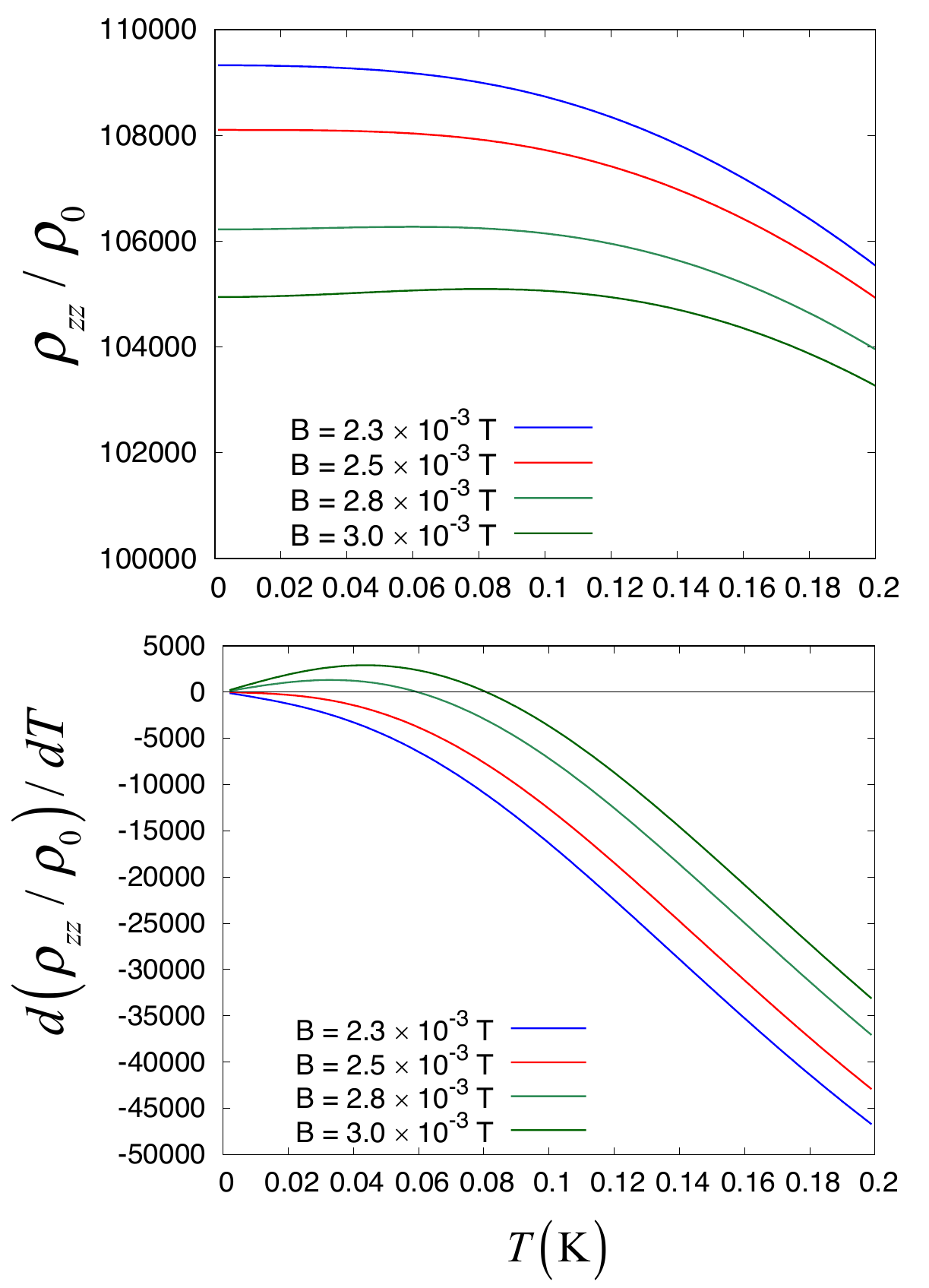}
  \caption{
    \label{fig:rzz_T_analysis}
    (Color online)
    The upper panel shows $\rho_{zz}/\rho_0$ as a function of $T$
    for different magnetic fields.
    There are peaks for $B>2.5\times 10^{-3}$ T
    but no peaks for $B<2.5\times 10^{-3}$ T.
    This behavior becomes clearer by taking
    the derivative of $\rho_{zz}$ with respect to $T$
    as shown in the lower panel.
  }
\end{figure}

Although it is not justified to extract information about
the Dirac fermion parameters from the finite temperature data,
it is possible to estimate $v_F/\Gamma$
from the critical value of $B=B_0$ that is obtained
by taking $T_{\rm max} \to 0$.
An analysis of $T_{\rm max}$ at small magnetic fields
is shown in Fig.~\ref{fig:rzz_T_analysis}.
We find $B=B_0=2.5\times 10^{-3}$~T
for the case of $v=5\times 10^4$ m/s and $\Gamma=1$ K.
From the value of $B_0$ we are able to evaluate $v_F/\Gamma$.
When the first Landau level energy is equal to $\Gamma$,
we obtain
\be
\frac{{{v_F}}}{\Gamma } = \frac{1}{{\sqrt {2\hbar eB} }}.
\ee
If we denote the Fermi velocity by ${v'_F}$
evaluated from this formula, assuming
$B=B_0=2.5\times 10^{-3}$~T
and $\Gamma=1$~K,
we find ${v'_F} = 0.95 \times {v_F}$.
The same analysis for $\Gamma=3$K 
leads to $B_0=2.3\times 10^{-2}$~T
and ${v'_F} = 0.94 \times {v_F}$.
Since $v'_F/v_F$ is very close to 1,
the value of $B_0$ can be used to evaluate $v_F/\Gamma$.

\subsection{Massive Dirac fermions}
\label{subsec:massive_case}
\begin{figure}[tbp]
  \includegraphics[width=\linewidth]{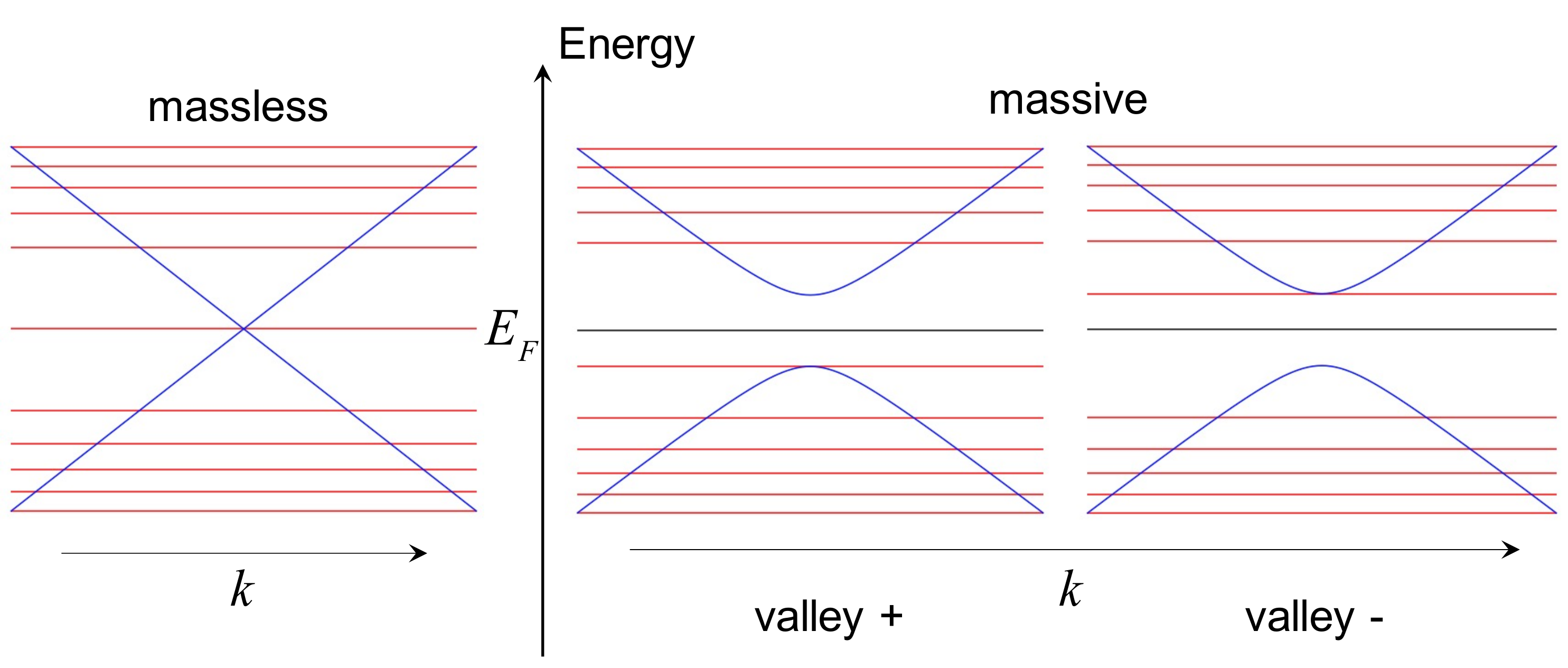}
  \caption{
    \label{fig:massive_spectrum}
    The energy dispersion of the massless case and the massive case.
      The Fermi energy is at the Dirac point.
      The horizontal lines are the Landau level energies.
      The zero energy Landau level is at $-m$ or $m$
      for the massive case.
      (The Zeeman energy is not included here for simplicity.)
  }
\end{figure}

Since \alphaI consists of light atoms,
the spin-orbit coupling is considered to be very small.
However, its value is estimated to be 1-2 meV based
on the density functional theory.\cite{Winter2017}
So, one may expect that massless Dirac fermions
become massive due to the presence of the spin-orbit coupling.\cite{Osada2018}
Since the Fermi energy is at the Dirac point in \alphaI,
  the mass gap affects the density of states at the Fermi energy.
  The energy spectrum of the massive Dirac fermions is schematically shown in Fig.~\ref{fig:massive_spectrum}.
  In the massless case, the negative magnetoresistance is associated with
  the appearance of the zero energy Landau levels at the Fermi energy.
  Meanwhile, this effect is suppressed for the massive case
  because the zero energy Landau level is located at $\pm m$
  for the massive case.
  Under magnetic fields, we also need to take into account the Zeeman energy,
  which is not shown in Fig.~\ref{fig:massive_spectrum}.
In the following we discuss the effect of the mass gap
on the interlayer resistivity under magnetic field.

\begin{figure}[tbp]
    \includegraphics[width=\linewidth]{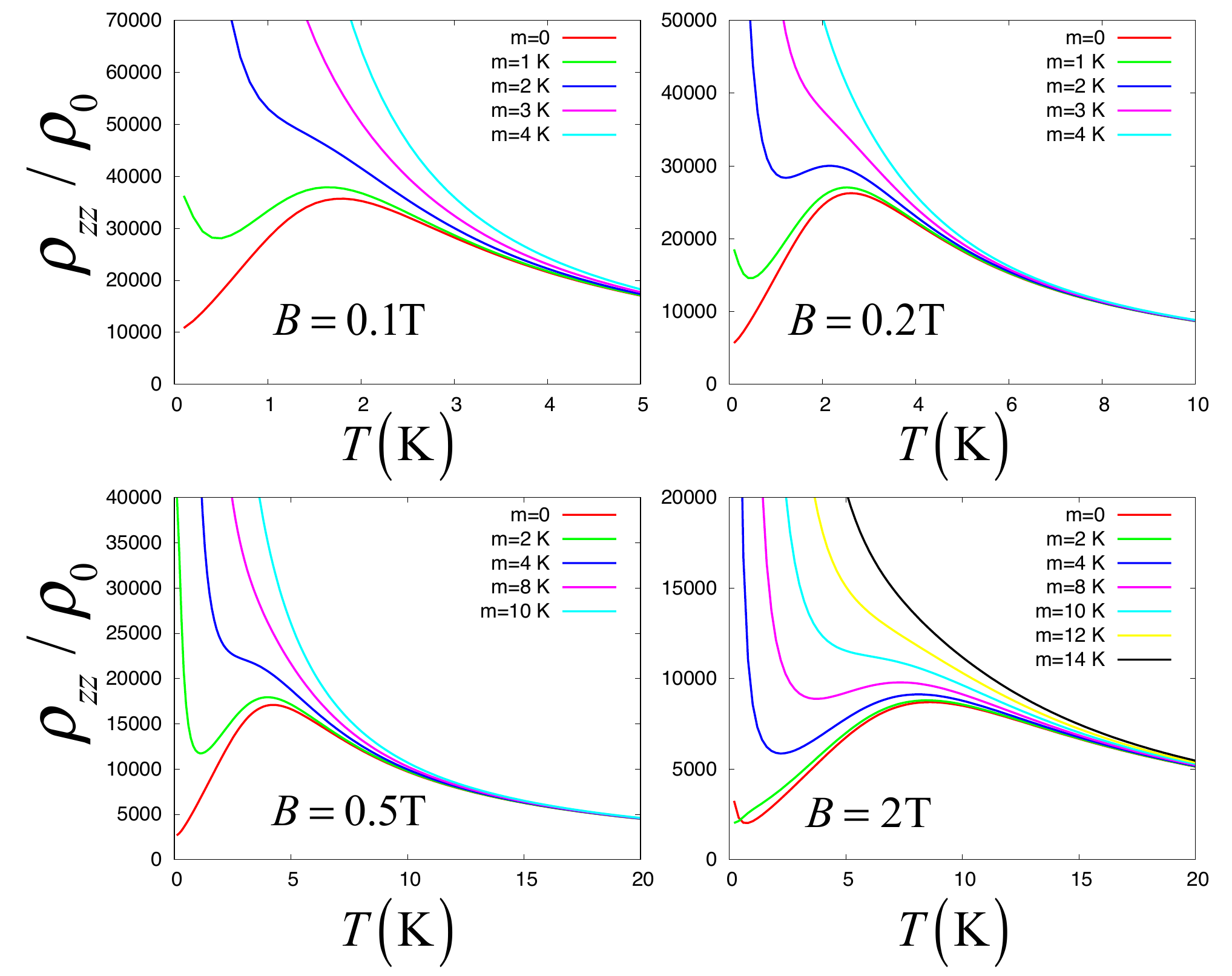}
  \caption{
    \label{fig:rzz_T_massive}
    (Color online)
    The interlayer resistivity as a function of temperature
    for different masses.
    From left top to right bottom,
    the magnetic field is $B=0.1, 0.2, 0.5, 2$T.
  }
\end{figure}

From the temperature dependence of the interlayer resistivity for different masses,
we find that the positive magnetoresistance region decreases as the mass increases, 
and finally disappear as shown in Fig. \ref{fig:rzz_T_massive}.
The critical mass value depends on $B$.
If the Dirac fermions are massive,
the positive magnetoresistance region disappears
at sufficiently low magnetic fields.

\begin{figure}[tbp]
  \includegraphics[width=\linewidth]{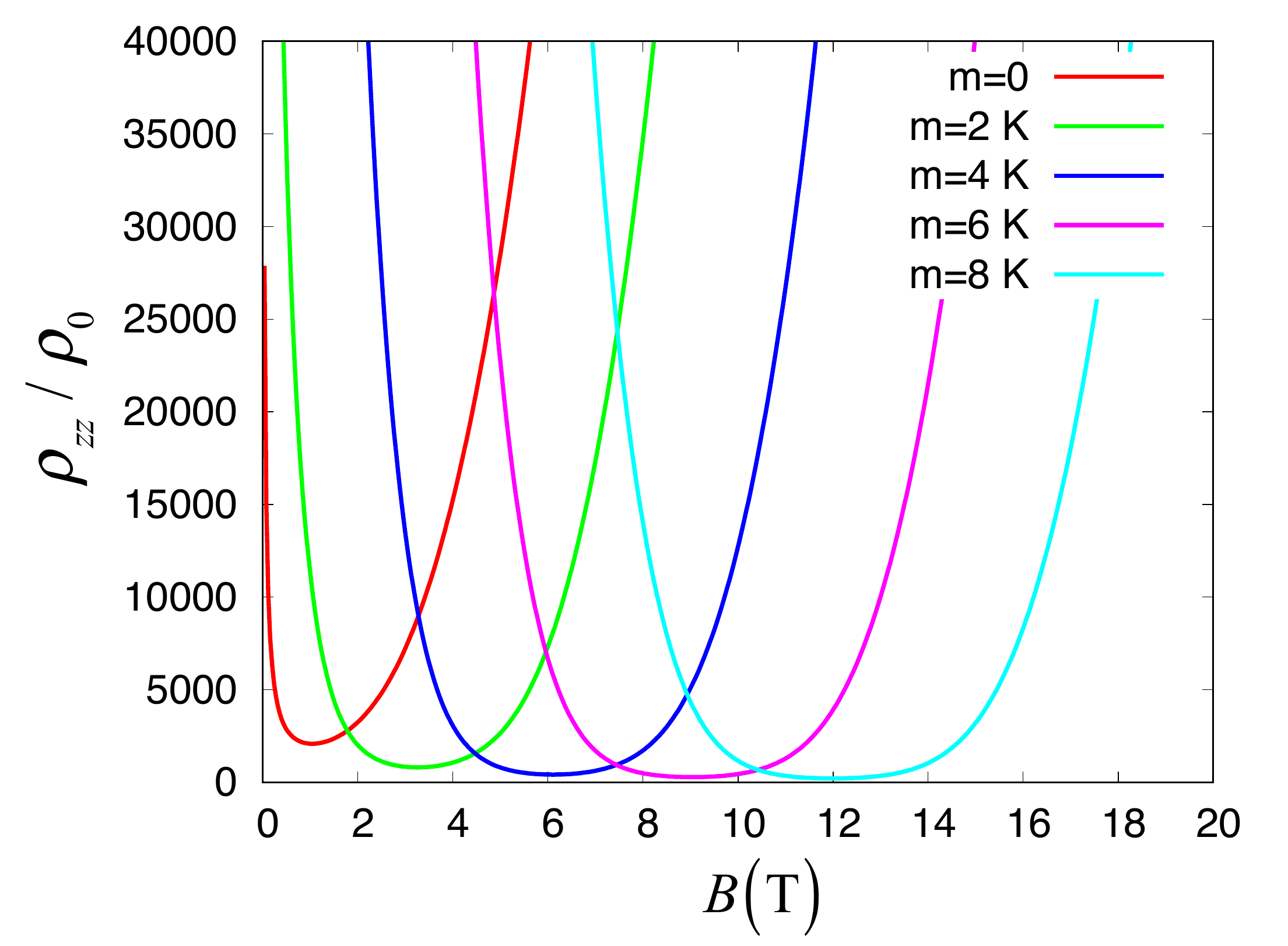}
  \caption{
    \label{fig:rzz_Bmin_massive}
    (Color online)
    The interlayer resistivity as a function of the magnetic field
    for different masses at $T=0.2$K.
    The value of the magnetic field at the resistivity minimum increases
    as the mass increases.
  }
\end{figure}

The effect of the mass term is clearly seen
at the minimum of the interlayer resistivity.
Figure \ref{fig:rzz_Bmin_massive} shows
the magnetic field dependence of the interlayer resistivity
for different masses in the magnetic field range where
the minimum is seen.
For the massless case the minimum is located where $\Gamma$
is equal to the Zeeman energy.\cite{Osada2008}
However, for massive cases, the minimum occurs at $\mu_{\rm B} B \simeq m$.
Thus, one can find the upper limit for the
mass term of the Dirac fermion system
from the minimum of the interlayer resistivity.
The massless case might appear different from Fig.~\ref{fig:rzz_for_Ts}.
  But we note that the temperature is 0.2~K and is much lower than Fig.~\ref{fig:rzz_for_Ts}.
  We also note that the minimum at $T=2K$ is located $B\sim 4$~T
  that is outside of the magnetic field range of Fig.~\ref{fig:rzz_for_Ts}.

\section{Conclusion}
\label{sec:conclusion}
To conclude, we have investigated the interlayer resistivity
under magnetic field with including
the intervalley scattering upon the interlayer tunneling.
The intervalley scattering occurs because of the presence
of the insulating layers.
The formula for the interlayer resistivity has been given
under magnetic field and for the case of zero-magnetic field.
From the numerical calculations
it is clearly seen that there is a crossover
from the positive to negative magnetoresistance regimes.
We have argued that the ratio of the Fermi velocity to the scattering rate,
$v_F/\Gamma$, can be estimated from the extrapolation of $B_0$
at finite temperatures to zero temperature.
The magnetic field dependence of the peaks
observed in the temperature dependence of the interlayer resistivity
is roughly proportional to $\sqrt{B}$.
However, one has to be careful about extracting information about
the Dirac fermion parameters from this kind of analysis
because the temperature effect is not simple.
As for the possibility of the mass gap in the Dirac fermion spectrum,
we have pointed out that the mass value can be estimated
from the minimum of the interlayer resistivity
as a function of the magnetic field.

\begin{acknowledgments}
  The author thanks N. Tajima for helpful discussions.
\end{acknowledgments}

\appendix
\section{Dirac Fermions in a Magnetic Field}
\label{app:DiracFermionLL}
In this appendix we briefly review the Dirac fermion Landau levels
with including a mass gap.
We consider Dirac fermions with the Fermi velocity $v_F$ and the mass gap $m$
at valley $\tau=+$
under magnetic field $B$, which is directed perpendicular
to the layer of the Dirac fermions.
The Hamiltonian is given by
\be
  {H_ + } = v_F\left( {\hbar {{\widehat k}_x} + e{A_x}} \right){\sigma _1}
  + v_F\left( {\hbar {{\widehat k}_y} + e{A_y}} \right){\sigma _2}
  + m{\sigma _3}.
\ee
Here,
${\widehat k_x} =  - i{\partial _x}$
and
${\widehat k_y} =  - i{\partial _y}$.
$\sigma_1$, $\sigma_2$, and $\sigma_3$ are Pauli matrices.
We use the Landau gauge, ${\bm{A}} = \left( { - By,0} \right)$.

The Landau level energy is given by
\be
  {\varepsilon _{n,+}} = \frac{{\sqrt 2 \hbar v_F}}{\ell }
  {\mathop{\rm sgn}} \left( n \right)\sqrt {\left| n \right| + {{\widetilde m}^2}},
  \ee
  for $n\neq 0$ and
  \be
    {\varepsilon _{0, + }} =  - m.
    \ee
    Here,
    $\ell  = \sqrt {\hbar /\left( {eB} \right)}$ is
    the magnetic length,
    $n$ is an integer,
    and
    $\widetilde m = m/\left( {\sqrt 2 \hbar v_F/\ell } \right)$.
  Denoting the Landau level states by
  $\left| {n, + } \right\rangle$,
  we obtain
  \be
  \left| {0, + } \right\rangle  = \left( {\begin{array}{*{20}{c}}
0\\
{\left| 0 \right\rangle }
  \end{array}} \right),
  \ee
  \be
  \left| {n, + } \right\rangle  = {C_n}\left( {\begin{array}{*{20}{c}}
      {\left( {\sqrt {n + {{\widetilde m}^2}}  + {\widetilde m}} \right)
        \left| {n - 1} \right\rangle }\\
{ - \sqrt n \left| n \right\rangle }
  \end{array}} \right),
  \ee
  for $n>0$, and
  \be
  \left| { n, + } \right\rangle  = {C_n}
    \left( {\begin{array}{*{20}{c}}
{\sqrt {\left| n \right|} \left| {\left| n \right| - 1} \right\rangle }\\
{\left( {\sqrt {\left| n \right| + {{\widetilde m}^2}}  + {\widetilde m}} \right)\left| {\left| n \right|} \right\rangle }
  \end{array}} \right)
  \ee
  for $n<0$.
  Here, 
  $\left| {\left| n \right|} \right\rangle $ are the harmonic oscillator
  eigenstates and
  \be
    {C_n} = \frac{1}{{\sqrt {2\sqrt {\left| n \right| + {{\widetilde m}^2}}
          \left( {\sqrt {\left| n \right| + {{\widetilde m}^2}}  + {\widetilde m}} \right)} }}.
  \ee
  
  The Landau levels at another valley, $\tau=-$,
  are obtained by making use of the transformation,
  \be
    {H_ - }\left( {\bm{k}} \right)
    = v_F\left( { - \hbar {k_x}{\sigma _1} + \hbar {k_y}{\sigma _2}} + m {\sigma_3} \right)
    = {\sigma _2}{H_ + }\left( {\bm{k}} \right){\sigma _2}.
    \ee
    The Landau level energies are
\be
  {\varepsilon _{n,-}} = {\varepsilon _{n,+}},
  \ee
  for $n \neq 0$ and
  \be
    {\varepsilon _{0, - }} = m.
    \ee
    The Landau level states are
    \be
    \left| {0, - } \right\rangle  = \left( {\begin{array}{*{20}{c}}
{\left| 0 \right\rangle }\\
0
    \end{array}} \right),
    \ee
    \be
    \left| {n, - } \right\rangle  = {C_n}\left( {\begin{array}{*{20}{c}}
{\left( {\sqrt {n + {{\widetilde m}^2}}  + \widetilde m} \right)\left| n \right\rangle }\\
{ - \sqrt n \left| {n - 1} \right\rangle }
    \end{array}} \right),
    \ee
    for $n>0$
    and
    \be
    \left| { n, - } \right\rangle  = {C_n}\left( {\begin{array}{*{20}{c}}
{\sqrt {\left| n \right|} \left| {\left| n \right|} \right\rangle }\\
{\left( {\sqrt {\left| n \right| + {{\widetilde m}^2}}  + \widetilde m} \right)\left| {\left| n \right| - 1} \right\rangle }
    \end{array}} \right),
    \ee
    for $n<0$.

\section{The Matrix Elements}
\label{app:M}
The matrix elements $M_{n\tau ,n'\tau '}$ are computed as follows.
For the intravalley tunneling, we simply obtain
\be
   {M_{n\tau ,n'\tau }} = {\delta _{n,n'}}.
\ee

In order to describe the intervalley tunneling,
we define
\be
   {A_n} = {C_n}\left( {\sqrt {n + { {\widetilde m}^2 }}  + {\widetilde m}} \right),
\ee             
and
\be
{B_n} = {C_n}\sqrt n.
\ee
In the following, $n$ and $n'$ denote non-negative integers.
In terms of $A_n$ and $B_n$,
we find
\bea
\left\langle {{0, + }}
 \mathrel{\left | {\vphantom {{0, + } {0, - }}}
 \right. \kern-\nulldelimiterspace}
 {{0, - }} \right\rangle  &=& 0, \nonumber \\
\left\langle {{0, + }}
 \mathrel{\left | {\vphantom {{0, + } {n, - }}}
 \right. \kern-\nulldelimiterspace}
 {{n, - }} \right\rangle  &=&  - {B_1}{\delta _{n,1}},
 \nonumber \\
 \left\langle {{0, + }}
 \mathrel{\left | {\vphantom {{0, + } { - n, - }}}
 \right. \kern-\nulldelimiterspace}
 {{ - n, - }} \right\rangle  &=& {A_1}{\delta _{n,1}},
 \nonumber \\
\left\langle {{n, + }}
 \mathrel{\left | {\vphantom {{n, + } {0, - }}}
 \right. \kern-\nulldelimiterspace}
         {{0, - }} \right\rangle  &=& {A_1}{\delta _{n,1}},
 \nonumber \\         
\left\langle {{n, + }}
 \mathrel{\left | {\vphantom {{n, + } {n', - }}}
 \right. \kern-\nulldelimiterspace}
         {{n', - }} \right\rangle
         &=& {\delta _{n,n' + 1}}{A_{n' + 1}}{A_{n'}}
         + {\delta _{n',n + 1}}{B_n}{B_{n + 1}},
         \nonumber \\
\left\langle {{n, + }}
 \mathrel{\left | {\vphantom {{n, + } { - n', - }}}
 \right. \kern-\nulldelimiterspace}
         {{ - n', - }} \right\rangle
         &=& {\delta _{n,n' + 1}}{A_{n' + 1}}{B_{n'}}
         - {\delta _{n',n + 1}}{B_n}{A_{n + 1}},
\nonumber \\
\left\langle {{ - n, + }}
 \mathrel{\left | {\vphantom {{ - n, + } {0, - }}}
 \right. \kern-\nulldelimiterspace}
 {{0, - }} \right\rangle &=& {B_1}{\delta _{n,1}},
\nonumber \\
\left\langle {{ - n, + }}
 \mathrel{\left | {\vphantom {{ - n, + } {n', - }}}
 \right. \kern-\nulldelimiterspace}
         {{n', - }} \right\rangle
         &=& {B_{n' + 1}}{A_{n'}}{\delta _{n,n' + 1}}
         - {A_n}{B_{n + 1}}{\delta _{n',n + 1}},
\nonumber \\
\left\langle {{ - n, + }}
 \mathrel{\left | {\vphantom {{ - n, + } { - n', - }}}
 \right. \kern-\nulldelimiterspace}
         {{ - n', - }} \right\rangle
         &=& {B_{n' + 1}}{B_{n'}}{\delta _{n,n' + 1}}
         + {A_n}{A_{n + 1}}{\delta _{n',n + 1}}.
         \nonumber \\
\eea

\section{Interlayer Conductivity in the Absence of Magnetic Field}
\label{app:B_0formula}
In this appendix we compute the interlayer conductivity
in the absence of the magnetic field.
We can use Eq.~(\ref{eq:s_zz_general})
with taking $\alpha  = \left( {{\bm{k}},\tau ,\sigma } \right)$
and $\alpha'  = \left( {{\bm{k}}',\tau' ,\sigma' } \right)$.
For $M_{\alpha '\alpha }$, we assume
\be
  {M_{\left( {{\bm{k'}},\tau ',\sigma '} \right),\left( {{\bm{k}},\tau ,\sigma } \right)}} = {\delta _{\sigma ',\sigma }}{\delta _{{\bm{k'}},{\bm{k}}}}.
  \ee
Note that the intervalley scattering is included.
  There is no spin flip upon interlayer tunneling.
  
  The ${\bm k}$-summation is converted into the integral:
  \bea
      {\sigma _{zz}} &=& \frac{{4\pi {e^2}t_c^2{a_c}}}{\hbar }\int {\frac{{{d^2}{\bm{k}}}}{{{{\left( {2\pi } \right)}^2}}}} \sum\limits_{s ,s' = \pm}
      \int_{ - \infty }^\infty  {d\omega } \left( { - \frac{{\partial f}}{{\partial \omega }}} \right) \nonumber \\
  & & \times {A_{{\bm{k}},s }}\left( \omega  \right)
        {A_{{\bm{k}},s'}}\left( \omega  \right),
    \eea
where
\be
   {A_{{\bm{k}},s }}\left( \omega  \right) =
   \frac{{\Gamma /\pi }}{
         {\left( \omega  -
               s \sqrt {{{\left( {\hbar {v_F}} \right)}^2}
                 \left( {k_x^2 + k_y^2} \right) + {m^2}}
             \right)^2 + {\Gamma ^2}
   }}.
  \ee
  Here, we introduced the same spectral broadening parameter $\Gamma$
  as in Eq.~(\ref{eq:s_zz_formula_B}).

The integration with respect to ${\bm k}$ is a standard one.
The result is,
\bea
    {\sigma _{zz}} &=& \frac{{2{e^2}t_c^2{a_c}}}{{{\pi ^2}{\hbar ^3}v_F^2}}\int_{ - \infty }^\infty  {d\omega } \left( { - \frac{{\partial f}}{{\partial \omega }}} \right)
\left[ 1 + \frac{1}{2}\left( {\frac{\omega }{\Gamma }
    + \frac{\Gamma }{\omega }} \right)
  \right. \nonumber \\
  & & \left. \times \left[ {{{\tan }^{ - 1}}\left( {\frac{{m + \omega }}{\Gamma }} \right) - {{\tan }^{ - 1}}\left( {\frac{{m - \omega }}{\Gamma }} \right)} \right]
     \right. \nonumber \\
     & & \left.
     + \frac{1}{2}\frac{{m\left( {\omega  - m} \right)}}{{{{\left( {\omega  - m} \right)}^2} + {\Gamma ^2}}} - \frac{1}{2}\frac{{m\left( {\omega  + m} \right)}}{{{{\left( {\omega  + m} \right)}^2} + {\Gamma ^2}}}
     \right].
  \eea
In case of massless Dirac fermions,
we obtain
\bea
    {\sigma _{zz}} &=& \frac{{2{e^2}t_c^2{a_c}}}{{{\pi ^2}{\hbar ^3}v_F^2}}\int_{ - \infty }^\infty  {d\omega } \left( { - \frac{{\partial f}}{{\partial \omega }}} \right)
    \nonumber \\
    & &  \times
    \left[ {1 + \left( {\frac{\omega }{\Gamma } + \frac{\Gamma }{\omega }} \right){{\tan }^{ - 1}}\left( {\frac{\omega }{\Gamma }} \right)} \right].
  \eea

\bibliography{../../../../refs/lib}

\begin{thebibliography}{17}%
\makeatletter
\providecommand \@ifxundefined [1]{%
 \@ifx{#1\undefined}
}%
\providecommand \@ifnum [1]{%
 \ifnum #1\expandafter \@firstoftwo
 \else \expandafter \@secondoftwo
 \fi
}%
\providecommand \@ifx [1]{%
 \ifx #1\expandafter \@firstoftwo
 \else \expandafter \@secondoftwo
 \fi
}%
\providecommand \natexlab [1]{#1}%
\providecommand \enquote  [1]{``#1''}%
\providecommand \bibnamefont  [1]{#1}%
\providecommand \bibfnamefont [1]{#1}%
\providecommand \citenamefont [1]{#1}%
\providecommand \href@noop [0]{\@secondoftwo}%
\providecommand \href [0]{\begingroup \@sanitize@url \@href}%
\providecommand \@href[1]{\@@startlink{#1}\@@href}%
\providecommand \@@href[1]{\endgroup#1\@@endlink}%
\providecommand \@sanitize@url [0]{\catcode `\\12\catcode `\$12\catcode
  `\&12\catcode `\#12\catcode `\^12\catcode `\_12\catcode `\%12\relax}%
\providecommand \@@startlink[1]{}%
\providecommand \@@endlink[0]{}%
\providecommand \url  [0]{\begingroup\@sanitize@url \@url }%
\providecommand \@url [1]{\endgroup\@href {#1}{\urlprefix }}%
\providecommand \urlprefix  [0]{URL }%
\providecommand \Eprint [0]{\href }%
\providecommand \doibase [0]{http://dx.doi.org/}%
\providecommand \selectlanguage [0]{\@gobble}%
\providecommand \bibinfo  [0]{\@secondoftwo}%
\providecommand \bibfield  [0]{\@secondoftwo}%
\providecommand \translation [1]{[#1]}%
\providecommand \BibitemOpen [0]{}%
\providecommand \bibitemStop [0]{}%
\providecommand \bibitemNoStop [0]{.\EOS\space}%
\providecommand \EOS [0]{\spacefactor3000\relax}%
\providecommand \BibitemShut  [1]{\csname bibitem#1\endcsname}%
\let\auto@bib@innerbib\@empty
\bibitem [{\citenamefont {Kobayashi}\ \emph {et~al.}(2004)\citenamefont
  {Kobayashi}, \citenamefont {Katayama}, \citenamefont {Noguchi},\ and\
  \citenamefont {Suzumura}}]{Kobayashi2004}%
  \BibitemOpen
  \bibfield  {author} {\bibinfo {author} {\bibfnamefont {A.}~\bibnamefont
  {Kobayashi}}, \bibinfo {author} {\bibfnamefont {S.}~\bibnamefont {Katayama}},
  \bibinfo {author} {\bibfnamefont {K.}~\bibnamefont {Noguchi}}, \ and\
  \bibinfo {author} {\bibfnamefont {Y.}~\bibnamefont {Suzumura}},\ }\href@noop
  {} {\bibfield  {journal} {\bibinfo  {journal} {J. Phys. Soc. Jpn.}\ }\textbf
  {\bibinfo {volume} {73}},\ \bibinfo {pages} {3135} (\bibinfo {year}
  {2004})}\BibitemShut {NoStop}%
\bibitem [{\citenamefont {Katayama}\ \emph {et~al.}(2006)\citenamefont
  {Katayama}, \citenamefont {Kobayashi},\ and\ \citenamefont
  {Suzumura}}]{Katayama2006}%
  \BibitemOpen
  \bibfield  {author} {\bibinfo {author} {\bibfnamefont {S.}~\bibnamefont
  {Katayama}}, \bibinfo {author} {\bibfnamefont {A.}~\bibnamefont {Kobayashi}},
  \ and\ \bibinfo {author} {\bibfnamefont {Y.}~\bibnamefont {Suzumura}},\
  }\href@noop {} {\bibfield  {journal} {\bibinfo  {journal} {J. Phys. Soc.
  Jpn.}\ }\textbf {\bibinfo {volume} {75}},\ \bibinfo {pages} {054705}
  (\bibinfo {year} {2006})}\BibitemShut {NoStop}%
\bibitem [{\citenamefont {Kajita}\ \emph {et~al.}(2014)\citenamefont {Kajita},
  \citenamefont {Nishio}, \citenamefont {Tajima}, \citenamefont {Suzumura},\
  and\ \citenamefont {Kobayashi}}]{Kajita2014}%
  \BibitemOpen
  \bibfield  {author} {\bibinfo {author} {\bibfnamefont {K.}~\bibnamefont
  {Kajita}}, \bibinfo {author} {\bibfnamefont {Y.}~\bibnamefont {Nishio}},
  \bibinfo {author} {\bibfnamefont {N.}~\bibnamefont {Tajima}}, \bibinfo
  {author} {\bibfnamefont {Y.}~\bibnamefont {Suzumura}}, \ and\ \bibinfo
  {author} {\bibfnamefont {A.}~\bibnamefont {Kobayashi}},\ }\href {\doibase
  10.7566/jpsj.83.072002} {\bibfield  {journal} {\bibinfo  {journal} {J. Phys.
  Soc. Jpn.}\ }\textbf {\bibinfo {volume} {83}},\ \bibinfo {pages} {072002}
  (\bibinfo {year} {2014})}\BibitemShut {NoStop}%
\bibitem [{\citenamefont {Unozawa}\ \emph {et~al.}(2020)\citenamefont
  {Unozawa}, \citenamefont {Kawasugi}, \citenamefont {Suda}, \citenamefont
  {Yamamoto}, \citenamefont {Kato}, \citenamefont {Nishio}, \citenamefont
  {Kajita}, \citenamefont {Morinari},\ and\ \citenamefont
  {Tajima}}]{Unozawa2020}%
  \BibitemOpen
  \bibfield  {author} {\bibinfo {author} {\bibfnamefont {Y.}~\bibnamefont
  {Unozawa}}, \bibinfo {author} {\bibfnamefont {Y.}~\bibnamefont {Kawasugi}},
  \bibinfo {author} {\bibfnamefont {M.}~\bibnamefont {Suda}}, \bibinfo {author}
  {\bibfnamefont {H.~M.}\ \bibnamefont {Yamamoto}}, \bibinfo {author}
  {\bibfnamefont {R.}~\bibnamefont {Kato}}, \bibinfo {author} {\bibfnamefont
  {Y.}~\bibnamefont {Nishio}}, \bibinfo {author} {\bibfnamefont
  {K.}~\bibnamefont {Kajita}}, \bibinfo {author} {\bibfnamefont
  {T.}~\bibnamefont {Morinari}}, \ and\ \bibinfo {author} {\bibfnamefont
  {N.}~\bibnamefont {Tajima}},\ }\href {\doibase 10.7566/jpsj.89.123702}
  {\bibfield  {journal} {\bibinfo  {journal} {J. Phys. Soc. Jpn.}\ }\textbf
  {\bibinfo {volume} {89}},\ \bibinfo {pages} {123702} (\bibinfo {year}
  {2020})}\BibitemShut {NoStop}%
\bibitem [{\citenamefont {Kobayashi}(2021)}]{Kobayashi2021}%
  \BibitemOpen
  \bibfield  {author} {\bibinfo {author} {\bibfnamefont {A.}~\bibnamefont
  {Kobayashi}},\ }\href {\doibase 10.7566/jpsjnc.18.02} {\bibfield  {journal}
  {\bibinfo  {journal} {{JPSJ} News and Comments}\ }\textbf {\bibinfo {volume}
  {18}},\ \bibinfo {pages} {02} (\bibinfo {year} {2021})}\BibitemShut {NoStop}%
\bibitem [{\citenamefont {Tajima}\ \emph {et~al.}(2009)\citenamefont {Tajima},
  \citenamefont {Sugawara}, \citenamefont {Kato}, \citenamefont {Nishio},\ and\
  \citenamefont {Kajita}}]{Tajima2009}%
  \BibitemOpen
  \bibfield  {author} {\bibinfo {author} {\bibfnamefont {N.}~\bibnamefont
  {Tajima}}, \bibinfo {author} {\bibfnamefont {S.}~\bibnamefont {Sugawara}},
  \bibinfo {author} {\bibfnamefont {R.}~\bibnamefont {Kato}}, \bibinfo {author}
  {\bibfnamefont {Y.}~\bibnamefont {Nishio}}, \ and\ \bibinfo {author}
  {\bibfnamefont {K.}~\bibnamefont {Kajita}},\ }\href@noop {} {\bibfield
  {journal} {\bibinfo  {journal} {Phys. Rev. Lett.}\ }\textbf {\bibinfo
  {volume} {102}},\ \bibinfo {pages} {176403} (\bibinfo {year}
  {2009})}\BibitemShut {NoStop}%
\bibitem [{\citenamefont {Osada}(2008)}]{Osada2008}%
  \BibitemOpen
  \bibfield  {author} {\bibinfo {author} {\bibfnamefont {T.}~\bibnamefont
  {Osada}},\ }\href@noop {} {\bibfield  {journal} {\bibinfo  {journal} {J.
  Phys. Soc. Jpn.}\ }\textbf {\bibinfo {volume} {77}},\ \bibinfo {pages}
  {084711} (\bibinfo {year} {2008})}\BibitemShut {NoStop}%
\bibitem [{\citenamefont {Morinari}\ and\ \citenamefont
  {Tohyama}(2010)}]{Morinari10a}%
  \BibitemOpen
  \bibfield  {author} {\bibinfo {author} {\bibfnamefont {T.}~\bibnamefont
  {Morinari}}\ and\ \bibinfo {author} {\bibfnamefont {T.}~\bibnamefont
  {Tohyama}},\ }\href@noop {} {\bibfield  {journal} {\bibinfo  {journal} {J.
  Phys. Soc. Jpn.}\ }\textbf {\bibinfo {volume} {79}},\ \bibinfo {pages}
  {044708} (\bibinfo {year} {2010})}\BibitemShut {NoStop}%
\bibitem [{\citenamefont {Winter}\ \emph {et~al.}(2017)\citenamefont {Winter},
  \citenamefont {Riedl},\ and\ \citenamefont {Valent{\'{\i}}}}]{Winter2017}%
  \BibitemOpen
  \bibfield  {author} {\bibinfo {author} {\bibfnamefont {S.~M.}\ \bibnamefont
  {Winter}}, \bibinfo {author} {\bibfnamefont {K.}~\bibnamefont {Riedl}}, \
  and\ \bibinfo {author} {\bibfnamefont {R.}~\bibnamefont {Valent{\'{\i}}}},\
  }\href {\doibase 10.1103/physrevb.95.060404} {\bibfield  {journal} {\bibinfo
  {journal} {Phys. Rev. B}\ }\textbf {\bibinfo {volume} {95}},\ \bibinfo
  {pages} {060404(R)} (\bibinfo {year} {2017})}\BibitemShut {NoStop}%
\bibitem [{\citenamefont {Osada}(2018)}]{Osada2018}%
  \BibitemOpen
  \bibfield  {author} {\bibinfo {author} {\bibfnamefont {T.}~\bibnamefont
  {Osada}},\ }\href {\doibase 10.7566/jpsj.87.075002} {\bibfield  {journal}
  {\bibinfo  {journal} {J. Phys. Soc. Jpn.}\ }\textbf {\bibinfo {volume}
  {87}},\ \bibinfo {pages} {075002} (\bibinfo {year} {2018})}\BibitemShut
  {NoStop}%
\bibitem [{\citenamefont {Kino}\ and\ \citenamefont
  {Miyazaki}(2006)}]{Kino2006}%
  \BibitemOpen
  \bibfield  {author} {\bibinfo {author} {\bibfnamefont {H.}~\bibnamefont
  {Kino}}\ and\ \bibinfo {author} {\bibfnamefont {T.}~\bibnamefont
  {Miyazaki}},\ }\href {\doibase 10.1143/jpsj.75.034704} {\bibfield  {journal}
  {\bibinfo  {journal} {J. Phys. Soc. Jpn.}\ }\textbf {\bibinfo {volume}
  {75}},\ \bibinfo {pages} {034704} (\bibinfo {year} {2006})}\BibitemShut
  {NoStop}%
\bibitem [{\citenamefont {McKenzie}\ and\ \citenamefont
  {Moses}(1998)}]{McKenzie1998}%
  \BibitemOpen
  \bibfield  {author} {\bibinfo {author} {\bibfnamefont {R.~H.}\ \bibnamefont
  {McKenzie}}\ and\ \bibinfo {author} {\bibfnamefont {P.}~\bibnamefont
  {Moses}},\ }\href {\doibase 10.1103/physrevlett.81.4492} {\bibfield
  {journal} {\bibinfo  {journal} {Phys. Rev. Lett.}\ }\textbf {\bibinfo
  {volume} {81}},\ \bibinfo {pages} {4492} (\bibinfo {year}
  {1998})}\BibitemShut {NoStop}%
\bibitem [{\citenamefont {Moses}\ and\ \citenamefont
  {McKenzie}(1999)}]{Moses1999}%
  \BibitemOpen
  \bibfield  {author} {\bibinfo {author} {\bibfnamefont {P.}~\bibnamefont
  {Moses}}\ and\ \bibinfo {author} {\bibfnamefont {R.~H.}\ \bibnamefont
  {McKenzie}},\ }\href {\doibase 10.1103/physrevb.60.7998} {\bibfield
  {journal} {\bibinfo  {journal} {Physical Review B}\ }\textbf {\bibinfo
  {volume} {60}},\ \bibinfo {pages} {7998} (\bibinfo {year}
  {1999})}\BibitemShut {NoStop}%
\bibitem [{\citenamefont {Mahan}(1990)}]{MahanChap9_3}%
  \BibitemOpen
  \bibfield  {author} {\bibinfo {author} {\bibfnamefont {G.~D.}\ \bibnamefont
  {Mahan}},\ }\enquote {\bibinfo {title} {Many-particle physics},}\ \ (\bibinfo
   {publisher} {Prenum, New York},\ \bibinfo {year} {1990})\ Chap.~\bibinfo
  {chapter} {9}, pp.\ \bibinfo {pages} {788--796},\ \bibinfo {edition} {2nd}\
  ed.\BibitemShut {Stop}%
\bibitem [{\citenamefont {Bender}\ \emph {et~al.}(1984)\citenamefont {Bender},
  \citenamefont {Hennig}, \citenamefont {Schweitzer}, \citenamefont {Dietz},
  \citenamefont {Endres},\ and\ \citenamefont {Keller}}]{Bender1984}%
  \BibitemOpen
  \bibfield  {author} {\bibinfo {author} {\bibfnamefont {K.}~\bibnamefont
  {Bender}}, \bibinfo {author} {\bibfnamefont {I.}~\bibnamefont {Hennig}},
  \bibinfo {author} {\bibfnamefont {D.}~\bibnamefont {Schweitzer}}, \bibinfo
  {author} {\bibfnamefont {K.}~\bibnamefont {Dietz}}, \bibinfo {author}
  {\bibfnamefont {H.}~\bibnamefont {Endres}}, \ and\ \bibinfo {author}
  {\bibfnamefont {H.~J.}\ \bibnamefont {Keller}},\ }\href@noop {} {\bibfield
  {journal} {\bibinfo  {journal} {Mol. Cryst. Liq. Cryst.}\ }\textbf {\bibinfo
  {volume} {108}},\ \bibinfo {pages} {359} (\bibinfo {year}
  {1984})}\BibitemShut {NoStop}%
\bibitem [{\citenamefont {Sugawara}\ \emph {et~al.}(2010)\citenamefont
  {Sugawara}, \citenamefont {Tamura}, \citenamefont {Tajima}, \citenamefont
  {Kato}, \citenamefont {Sato}, \citenamefont {Nishio},\ and\ \citenamefont
  {Kajita}}]{Sugawara2010mr}%
  \BibitemOpen
  \bibfield  {author} {\bibinfo {author} {\bibfnamefont {S.}~\bibnamefont
  {Sugawara}}, \bibinfo {author} {\bibfnamefont {M.}~\bibnamefont {Tamura}},
  \bibinfo {author} {\bibfnamefont {N.}~\bibnamefont {Tajima}}, \bibinfo
  {author} {\bibfnamefont {R.}~\bibnamefont {Kato}}, \bibinfo {author}
  {\bibfnamefont {M.}~\bibnamefont {Sato}}, \bibinfo {author} {\bibfnamefont
  {Y.}~\bibnamefont {Nishio}}, \ and\ \bibinfo {author} {\bibfnamefont
  {K.}~\bibnamefont {Kajita}},\ }\href {\doibase 10.1143/JPSJ.79.113704}
  {\bibfield  {journal} {\bibinfo  {journal} {J. Phys. Soc. Jpn.}\ }\textbf
  {\bibinfo {volume} {79}},\ \bibinfo {pages} {113704} (\bibinfo {year}
  {2010})}\BibitemShut {NoStop}%
\bibitem [{\citenamefont {Yoshimura}\ \emph {et~al.}(2021)\citenamefont
  {Yoshimura}, \citenamefont {Sato},\ and\ \citenamefont
  {Osada}}]{Yoshimura2021}%
  \BibitemOpen
  \bibfield  {author} {\bibinfo {author} {\bibfnamefont {K.}~\bibnamefont
  {Yoshimura}}, \bibinfo {author} {\bibfnamefont {M.}~\bibnamefont {Sato}}, \
  and\ \bibinfo {author} {\bibfnamefont {T.}~\bibnamefont {Osada}},\ }\href
  {\doibase 10.7566/jpsj.90.033701} {\bibfield  {journal} {\bibinfo  {journal}
  {J. Phys. Soc. Jpn.}\ }\textbf {\bibinfo {volume} {90}},\ \bibinfo {pages}
  {033701} (\bibinfo {year} {2021})}\BibitemShut {NoStop}%
\end{thebibliography}%
\end{document}